\documentclass[aps,prd,onecolumn,showpacs,groupedaddress, nofootinbib]{revtex4}  
\usepackage{graphicx}
\usepackage{epstopdf}
\usepackage{amsmath}
\usepackage{amsfonts}
\usepackage{amssymb}
\usepackage{appendix}
\usepackage{comment}
\usepackage{bbold}
\usepackage{color}
\usepackage{slashed}
\usepackage{subfigure}
\usepackage{setspace}
\usepackage{footnote}
\usepackage{multirow}

\usepackage{url}
\usepackage{wrapfig}
\usepackage{braket}

\begin{document}
\singlespacing
{\hfill NUHEP-TH/15-04}

\title{Non-standard Neutrino Interactions at DUNE}

\author{Andr\'{e} de Gouv\^{e}a}
\author{Kevin J. Kelly}
\affiliation{Northwestern University, Department of Physics \& Astronomy, 2145 Sheridan Road, Evanston, IL 60208, USA}

\begin{abstract}
We explore the effects of non-standard neutrino interactions (NSI) and how they modify neutrino propagation in the Deep Underground Neutrino Experiment (DUNE). We find that NSI can significantly modify the data to be collected by the DUNE  experiment as long as the new physics parameters are large enough. For example, If the DUNE data are consistent with the standard three-massive-neutrinos paradigm, order 0.1 (in units of the Fermi constant) NSI effects will be ruled out. On the other hand, if large NSI effects are present, DUNE will be able to not only rule out the standard paradigm but also measure the new physics parameters, sometimes with good precision. We find that, in some cases, DUNE is sensitive to new sources of $CP$-invariance violation. We also explored whether DUNE data can be used to distinguish different types of new physics beyond nonzero neutrino masses. In more detail, we asked whether NSI can be mimicked, as far as the DUNE setup is concerned, by the hypothesis that there is a new light neutrino state.
\end{abstract}

\maketitle

\section{Introduction}
\label{sec:Introduction}

The main goals of next-generation long-baseline neutrino experiments like the Deep Underground Neutrino Experiment (DUNE) (see, for example, Ref.~\cite{Adams:2013qkq}) and Hyper-Kamiokande (HyperK) \cite{Abe:2015zbg} are to search for leptonic $CP$-invariance violation and to test the three-massive-neutrinos paradigm (standard paradigm). 

The standard paradigm consists of postulating that the three active neutrinos $\nu_{e,\mu,\tau}$ are linear combination of three neutrino states with well-defined mass $\nu_{1,2,3}$ (masses $m_{1,2,3}$, respectively), and that the mechanism behind the nonzero neutrino masses is such that no accessible interactions or states beyond those prescribed in the standard model of particle physics (SM) are present. In summary, neutrino production, detection, and propagation (including matter effects) are described by the weak interactions, and there are no new light degrees of freedom. While all neutrino data collected so far -- with the exception of the short-baseline anomalies \cite{Aguilar:2001ty,AguilarArevalo:2008rc,Mention:2011rk,Frekers:2011zz,Aguilar-Arevalo:2013pmq}, which we will not consider here -- are consistent with the standard paradigm, large deviations are allowed. Candidates for the new physics beyond the standard paradigm include more than three light neutrino mass-eigenstates, new ``weaker-than-weak'' neutrino--matter interactions, small violations of fundamental physics principles, and couplings between neutrinos and new very light states. 

One way to test the standard paradigm is to measure, as precisely as possible, what is interpreted to be the neutrino oscillation probabilities ($\nu_{\mu}\to\nu_{\mu}, \nu_{\mu}\to\nu_{e}, \bar{\nu}_{\mu}\to\bar{\nu}_{\mu}$, etc) as a function of the baseline $L$ or the neutrino energy $E_{\nu}$, and compare the results  with the expectations of the standard paradigm. If the observed pattern is not consistent with expectations, one can conclude that there is new physics beyond nonzero neutrino masses. 

Here we investigate how well precision measurements of oscillation probabilities at DUNE can be used to probe the existence of non-standard neutrino neutral-current-like interactions (NSI). In more detail, we discuss how much DUNE data, assuming it is consistent with the standard paradigm, can constrain NSI, and discuss, assuming NSI are present, how well DUNE can measure the new physics parameters, including new sources of $CP$-invariance violation. We also investigate whether DUNE has the ability to differentiate between different types of new phenomena by comparing NSI effects with those expected from the existence of a new neutrino-like state (sterile neutrino).  

Studies of the effects of new neutrino--matter interactions on neutrino propagation are not new. Indeed, they were first proposed in Wolfenstein's seminal paper that introduced matter effects \cite{Wolfenstein:1977ue}. NSI were explored as a solution to the solar neutrino problem  \cite{Guzzo:1991hi}, and their impact on the oscillations of solar neutrinos \cite{Krastev:1992zx,Miranda:2004nb,Bolanos:2008km,Palazzo:2009rb,Escrihuela:2009up,Friedland:2004pp}, atmospheric neutrinos \cite{GonzalezGarcia:1998hj,Fornengo:1999zp,Fornengo:2001pm,Huber:2001zw,Friedland:2004ah,Friedland:2005vy,Yasuda:2010hw,GonzalezGarcia:2011my,Esmaili:2013fva,Choubey:2014iia,Mocioiu:2014gua,Fukasawa:2015jaa,Choubey:2015xha}, and accelerator neutrinos \cite{Friedland:2006pi,Blennow:2007pu,EstebanPretel:2008qi,Kopp:2010qt,Coloma:2011rq,Friedland:2012tq,Coelho:2012bp,Adamson:2013ovz,Girardi:2014kca,Blennow:2015nxa} have been thoroughly explored in the literature in the last twenty years. Some consequences of NSI to DUNE were also explored very recently in Ref.~\cite{Masud:2015xva}. We add to the discussion in several ways. We consider all relevant NSI parameters when estimating the reach of DUNE, and investigate the capabilities of the experiment to see the new sources of $CP$-invariance violation. We also perform a detailed simulation of the experiment and deal with the concurrent measurements of the NSI parameters and the standard oscillation parameters. In addition, we address whether and how next-generation long-baseline experiments can distinguish different manifestations of new physics (other than nonzero neutrino masses) in the lepton sector. 

This manuscript is organized as follows. In Sec.~\ref{sec:Formalism}, we discuss NSI and their effects on neutrino oscillations via nonstandard matter effects. We also review, briefly, what is currently known about NSI and discuss some of the assumptions we make implicitly and explicitly. In Sec.~\ref{sec:Exclusion}, we briefly discuss some of the details of our simulations of the DUNE experiment and compute how well DUNE can exclude the various NSI new physics parameters. In Sec.~\ref{sec:Sensitivity}, we choose three different NSI scenarios and compute how well DUNE can measure the new physics parameters. Here we also address whether DUNE can distinguish between the existence of NSI and sterile neutrinos. In Sec.~\ref{sec:Conclusions}, we qualitatively discuss potential diagnostic tools beyond DUNE, including future data from HyperK and next-generation measurements of the atmospheric neutrino flux, and offer some concluding remarks.

\section{Formalism, Current Bounds, and Oscillation Probabilities}
\label{sec:Formalism}

We allow for the existence of new neutrino--matter interactions which, at low enough energies and after electroweak symmetry breaking, can be expressed in terms of dimension-six four-fermion interactions. In more detail, we consider the effective Lagrangian,
\begin{equation}
\mathcal{L}^{\mathrm{NSI}} = -2\sqrt{2}G_F (\bar{\nu}_{\alpha}\gamma_\rho \nu_\beta)(\epsilon_{\alpha\beta}^{f\widetilde{f}L} \overline{f}_L \gamma^\rho \widetilde{f}_L + \epsilon_{\alpha\beta}^{f\widetilde{f}R} \overline{f}_R \gamma^\rho \widetilde{f}_R) + h.c.,
\label{eq:L}
\end{equation}
where $G_F$ is the Fermi constant and $\epsilon_{\alpha\beta}^{f\widetilde{f} s}$ represent the interaction strength, relative to low-energy weak interactions, between neutrinos of flavor $\alpha$ and $\beta$, $\alpha,\beta=e,\mu,\tau$, with fermions $f_s$ and $\widetilde{f_s}$ of chirality $s$. We are only interested in the couplings to first generation charged fermions, $f=e,u,d$, and for simplicity only consider the diagonal couplings to the charged fermions, i.e., $f=\widetilde{f}$. While non-standard neutrino interactions may affect the neutrino production and detection processes, here we only consider the effects of the new interactions during neutrino propagation, as including these effects requires additional parameters describing the production and detection processes~(see, for example, Refs.~\cite{Huber:2002bi,Huber:2010dx}). The use of a near detector in experiments will help address these effects (see, for example, Ref.~\cite{Antusch:2010fe}. See also \cite{Khan:2013hva,Khan:2014zwa}). Following the standard conventions from Refs.~\citep{Ohlsson:2012kf,Choubey:2014iia,GonzalezGarcia:2011my,Kikuchi:2008vq,Yasuda:2010hw,Friedland:2006pi,Friedland:2005vy,Friedland:2004ah,Friedland:2004pp}, oscillation probabilities will be expressed in terms of the effective parameters $\epsilon_{\alpha\beta} = \sum_{f = u, d, e} \epsilon_{\alpha\beta}^f \frac{n_f}{n_e}$, where $\epsilon_{\alpha\beta}^f \equiv \epsilon_{\alpha\beta}^{f L} + \epsilon_{\alpha\beta}^{f R}$, $\epsilon_{\alpha\beta}^{f s} \equiv \epsilon_{\alpha\beta}^{f f s}$ and $n_f$ is the number density of fermion $f$. In the Earth's crust, assuming equal numbers of electrons, protons and neutrons, $n_u/n_e = n_d/n_e = 3$.

The probability for a neutrino flavor eigenstate $\ket{\nu_\alpha}$ to be detected as an eigenstate $\ket{\nu_\beta}$ is $P_{\alpha\beta} = |\mathcal{A}_{\alpha\beta}|^2$, where $\mathcal{A}_{\alpha\beta}$ is the oscillation amplitude. We can write
\begin{equation}
\mathcal{A}_{\alpha\beta} = \bra{\nu_\beta} U e^{-i H_{ij} L} U^\dagger \ket{\nu_\alpha},
\end{equation}
where $L$ is the propagation length of the neutrinos (assumed to be ultra-relativistic), $U$ is the Pontecorvo-Maki-Nakagawa-Sakata (PMNS) leptonic mixing matrix, and $H_{ij}$ is the propagation Hamiltonian in the basis of the neutrino mass eigenstates, which will be defined later, cf.~Eq.~(\ref{eq:Hij}). 

The PMNS matrix can be parameterized by three angles ($\theta_{ij}$; $i < j$; $i$, $j$ = 1, 2, 3) and one $CP$-violating phase ($\delta$). Using the standard parameterization from the particle data group \cite{Agashe:2014kda},
\begin{equation}
U = \left(\begin{array}{ccc}
c_{12} c_{13} & s_{12} c_{13} & s_{13}e^{-i\delta} \\
-s_{12} c_{13} - c_{12} s_{13} s_{23} e^{i\delta} & c_{12} c_{23} - s_{12} s_{13} s_{23} e^{i\delta} & c_{13} s_{23} \\
s_{12} s_{23} - c_{12} s_{13} c_{23} e^{i\delta} & -c_{12} s_{23} - s_{12} s_{13} c_{23} e^{i\delta} & c_{13} c_{23} \end{array}\right),
\end{equation}
where $c_{ij} = \cos{\theta_{ij}}$ and $s_{ij} = \sin{\theta_{ij}}$. The Hamiltonian is
\begin{equation}
H_{ij} = \frac{1}{2 E_\nu} \mathrm{diag}\left\lbrace 0, \Delta m_{12}^2, \Delta m_{13}^2\right\rbrace + V_{ij},
\label{eq:Hij}
\end{equation}
where $\Delta m_{ij}^2 = m_j^2 - m_i^2$, $E_\nu$ is the neutrino energy, and $V_{ij}$ is the matter potential which includes both the charged-current interactions with electrons and the NSI. As usual, SM neutral-current interactions can be absorbed as an overall phase in the propagation. Here, the matter potential is 
\begin{eqnarray}
V_{ij} &=& U^\dagger_{i\alpha} V_{\alpha\beta} U_{\beta j}, \\
V_{\alpha\beta} &=& A\left( \begin{array}{ccc}
1+\epsilon_{ee} & \epsilon_{e\mu} & \epsilon_{e\tau} \\
\epsilon_{e\mu}^* & \epsilon_{\mu\mu} & \epsilon_{\mu\tau} \\
\epsilon_{e\tau}^* & \epsilon_{\mu\tau}^* & \epsilon_{\tau\tau} \end{array} \right),
\end{eqnarray}
where $A = \sqrt{2} G_F n_e$ and $n_e$ is the number density of electrons along the path of propagation. For propagation through the Earth's crust, $A \simeq 10^{-4} \mathrm{\ eV}^2/\mathrm{GeV}$. The NSI parameters $\epsilon_{\alpha\beta}$ are real for $\alpha = \beta$ and complex for $\alpha \neq \beta$, so there are a total of nine new real parameters. For the purposes of oscillation probabilities, however, one is free to set any one of the diagonal $\epsilon_{\alpha\alpha}$ to zero. In other words, one can only constrain two independent differences of $\epsilon_{\alpha\alpha}$, e.g., $\epsilon_{ee}-\epsilon_{\mu\mu}$ and $\epsilon_{\tau\tau}-\epsilon_{\mu\mu}$. In calculating the oscillation probabilities for antineutrinos\footnote{We denote oscillation probabilities among antineutrinos as $P_{\bar{\alpha}\bar{\beta}}$.}, $U^*$ replaces $U$, and $-V_{ij}^{*}=-V_{ij}^T$ replaces $V_{ij}$, keeping in mind that $V$, which is a term in the neutrino propagation Hamiltonian, is Hermitian.

Data from neutrino oscillation experiments and other experimental probes are consistent with $\epsilon_{\alpha\beta} = 0$, and the most conservative, ``neutrino-only'' limits are, according to Ref.~\cite{Ohlsson:2012kf}, 
\begin{equation}
\left( \begin{array}{ccc}
|\epsilon_{ee}| < 4.2 & |\epsilon_{e\mu}| < 0.33 & |\epsilon_{e\tau}| < 3.0 \\
 & |\epsilon_{\mu\mu}| < 0.07 & |\epsilon_{\mu\tau}| < 0.33 \\
 &  & |\epsilon_{\tau\tau}| < 21 \end{array} \right),\label{Bounds}
\end{equation}
 and nothing is known about the phases of $\epsilon_{e\mu}$, $\epsilon_{e\tau}$, and $\epsilon_{\mu\tau}$~\cite{Miranda:2015dra,Ohlsson:2012kf,Biggio:2009kv,Davidson:2003ha}. We refer readers to the literature for all the details and provide some comments on these limits and the assumptions behind the effective theory Eq.~(\ref{eq:L}) in Sec.~\ref{sec:Conclusions}. Unless otherwise noted, we will use Eq.~(\ref{Bounds}) in order to gauge the reach of future oscillation experiments. 
  
For illustrative purposes, we consider a two-neutrino scheme with oscillations among $\nu_\mu$ and $\nu_\tau$,\footnote{This turns out to be a good approximation in the limit $\Delta m^2_{12}\to 0$ and $\sin^2\theta_{13}\to 0$. For this reason we use the suggestive labels $\Delta m^2_{13}$ and $\theta_{23}$ for the mass-squared difference and mixing angle, respectively.} allowing for either diagonal ($\epsilon_{\tau\tau}$) or off-diagonal ($\epsilon_{\mu\tau}$) NSI, and compute the muon-neutrino survival probabilities $P_{\mu\mu}$. In the case of only diagonal NSI ($\epsilon_{\tau\tau}\neq 0$),
\begin{equation}
\label{Diag}
P_{\mu\mu} = 1 - \frac{1}{1+\zeta}\sin^2{(2\theta_{23})}\sin^2{\left(\frac{\Delta_{13}}{2}\sqrt{1+\zeta}\right)},
\end{equation}
where $\Delta_{13} \equiv \Delta m_{13}^2 L/2E_\nu$, $L$ is oscillation baseline, $E_\nu$ is the neutrino energy, and $\zeta \equiv 2\eta_D \cos{(2\theta_{23})} + \eta_D^2$, with $\eta_D \equiv AE_\nu \epsilon_{\tau\tau}/\Delta m_{13}^2$. Nontrivial $\epsilon_{\tau\tau}$ (i.e., nonzero $\zeta$) modify the oscillation frequency and amplitude in matter relative to oscillations in vacuum. For neutrino energies $\mathcal{O}(1\mathrm{\ GeV})$ and $\Delta m_{13}^2 \simeq 10^{-3}$ eV$^2$, $\eta_D$ is a small parameter [$\mathcal{O}(0.1)$] even for $\epsilon_{\tau\tau}$ values around unity. Note that this oscillation probability is identical, once one reinterprets the mass-squared difference and the mixing angle, to two-flavor $\nu_e\to\nu_e$ oscillations in the presence of constant matter when SM matter effects are taken into account. 

For purely off-diagonal NSI ($\epsilon_{\mu\tau}\neq 0$), the disappearance probability is
\begin{equation}
\label{OffDiag}
P_{\mu\mu} = 1-\frac{1}{1+\xi}\left(\sin^2{(2\theta_{23})}+\xi\right)\sin^2{\left(\frac{\Delta_{13}}{2}\sqrt{1+\xi}\right)},
\end{equation}
where $\xi \equiv 4\Re{(\eta_O)}\sin{(2\theta_{23})}+4|\eta_O|^2$, and $\eta_O \equiv AE_\nu \epsilon_{\mu\tau}/\Delta m_{13}^2$. Again, $\eta_O$ is a small parameter for the neutrino energies of interest. General two-flavor oscillation probabilities in the presence of constant matter can be found, for example, in Ref.~\cite{DeGouvea:2001ag}.

Comparing Eqs.~(\ref{Diag}) and Eq.~(\ref{OffDiag}), it is easy to see that diagonal and non-diagonal NSI are qualitatively different. In the limit of large NSI ($\epsilon\to\infty$), for example, $P_{\mu\mu} \to 1$ in Eq.~(\ref{Diag}), because the muon-neutrino state becomes an eigenstate of the Hamiltonian. In the same limit,  $P_{\mu\mu} \to 1/2$ (assuming the oscillatory term averages out) in Eq.~(\ref{OffDiag}), i.e, the effective mixing is maximal. This occurs because, in the purely off-diagonal case, large NSI limit, the propagation Hamiltonian is purely off-diagonal. On the other hand, in the small $L$ limit ($\Delta_{13}\to 0$), $P_{\mu\mu}$ behaves as if the oscillations were taking place in vacuum in Eq.~(\ref{Diag}), while this is not the case in  Eq.~(\ref{OffDiag}). In the off-diagonal case, in the small $L$ limit, the ``matter mixing angle'' is $\sin^22\theta_{\rm eff}\simeq\sin^22\theta_{23}+\xi$.  We also note that $P_{\mu\mu}$ in Eq.~(\ref{OffDiag}) depends on $\Re{(\eta_O)}$ and $|\eta_O|^2$, and that, for a pure-imaginary $\epsilon_{\mu\tau}=i|\epsilon_{\mu\tau}|$, the NSI effects are proportional to $|\epsilon_{\mu\tau}|^2$. Therefore, for small $\epsilon_{\mu\tau}$, we expect lower sensitivity to NSI if the new-physics parameters are pure-imaginary.

\begin{figure}[ht]
\centering
\includegraphics[width=0.45\textwidth]{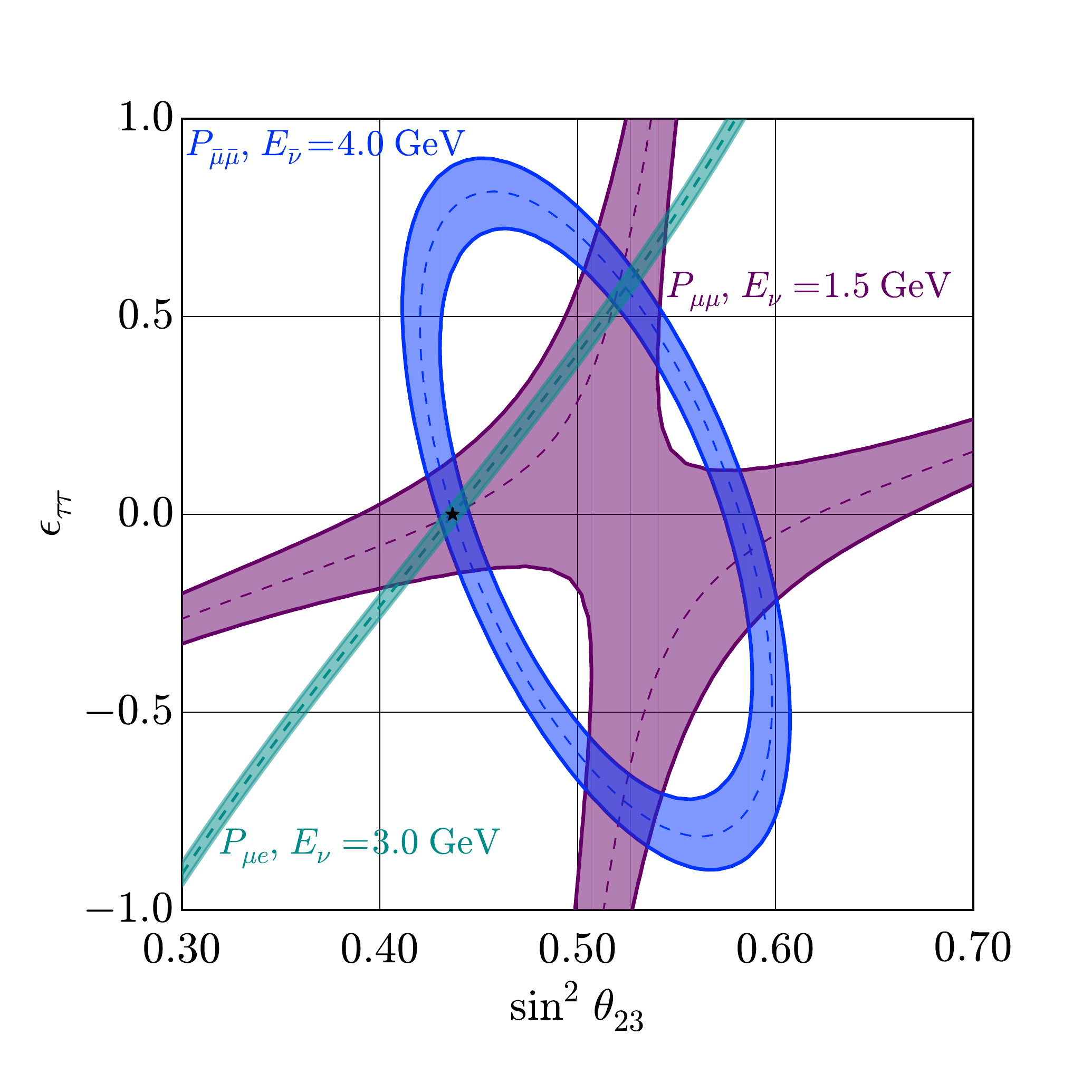}
\caption{Degeneracies -- points in the parameter space where the oscillation probabilities are the same for a fixed neutrino energy -- in the $\sin^2\theta_{23}$ and $\epsilon_{\tau\tau}$ parameter space for neutrino appearance (teal), neutrino disappearance (purple), and antineutrino disappearance (blue) at listed fixed energy, for $L=1300$ km. Dotted lines indicate exact degeneracy, while shaded regions indicate $\pm 1\%$ differences, or smaller (in magnitude). Three-neutrino oscillation parameters used for comparison are $\sin^2\theta_{12} = 0.308,\ \sin^2\theta_{13} = 0.0234,\ \sin^2\theta_{23} = 0.437,\ \Delta m_{12}^2 = 7.54 \times 10^{-5}$ eV$^2$, $\Delta m_{13}^2 = +2.47\times 10^{-3}$ eV$^2$ (in agreement with Ref.~\cite{Agashe:2014kda}) and $\delta = 0$.}
\label{fig:Degeneracy}
\end{figure}

Previous studies have found degeneracies in oscillation probabilities among different NSI parameters, as well as between NSI parameters and three-neutrino oscillation parameters~\cite{Friedland:2004ah}. In particular, these degeneracies arise when certain approximations are valid, e.g., $\sin^2\theta_{13} \to 0$ and $\Delta m_{12}^2 \to 0$ for atmospheric neutrino searches. We find that, for a long-baseline experiment, exact degeneracies exist for fixed values of $E_\nu$, but these are mostly broken when multiple energy bins and oscillation channels are combined. Fig.~\ref{fig:Degeneracy} depicts the degeneracy between $\sin^2\theta_{23}$ and $\epsilon_{\tau\tau}$ for fixed neutrino energy values for the muon-neutrino disappearance, muon-antineutrino disappearance, and the $\nu_{\mu}\to\nu_e$ neutrino appearance channels computed using a full three-flavor hypothesis. The fixed neutrino energy values are chosen to coincide with the maximum yields at DUNE, explained in more detail in Sec.~\ref{sec:Exclusion}. Three-neutrino oscillation parameters used for comparison are $\sin^2\theta_{12} = 0.308,\ \sin^2\theta_{13} = 0.0234,\ \sin^2\theta_{23} = 0.437,\ \Delta m_{12}^2 = 7.54 \times 10^{-5}$ eV$^2$, $\Delta m_{13}^2 = +2.47\times 10^{-3}$ eV$^2$. These parameters are in agreement with Ref.~\cite{Agashe:2014kda}, and we will refer to them in the following sections. Additionally, here, we set $\delta = 0$. Curves for the disappearance channels can be understood using the simplified form of the oscillation probability Eq.~(\ref{Diag}), and are nearly identical to those in Fig.~\ref{fig:Degeneracy}. The three curves in Fig.~\ref{fig:Degeneracy} also nearly intersect for $(\sin^2\theta_{23}, \epsilon_{\tau\tau}) \simeq (0.52, 0.6)$. In the next Sections, we will see manifestations of this phenomenon. 

\begin{figure}[htbp]
\centering
\includegraphics[width=0.6\linewidth]{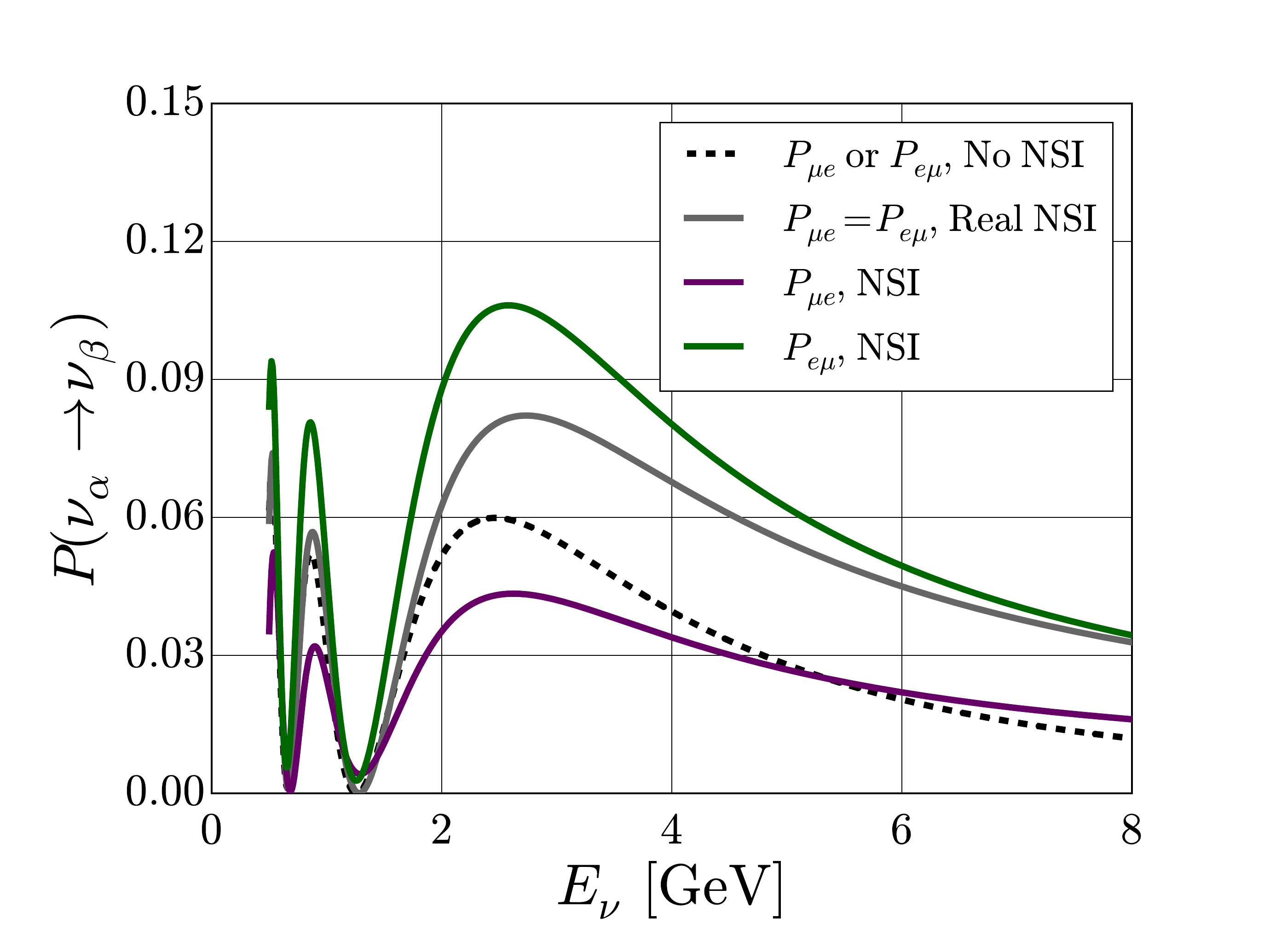}
\caption{$T$-invariance violating effects of NSI at $L=1300$~km for $\epsilon_{e\mu} = 0.1e^{i\pi/3}$, $\epsilon_{e\tau} = 0.1e^{-i\pi/4}$, $\epsilon_{\mu\tau} = 0.1$ (all other NSI parameters are set to zero). Here, the three-neutrino oscillation parameters are $\sin^2\theta_{12} = 0.308$, $\sin^2\theta_{13} = 0.0234$, $\sin^2\theta_{23} = 0.437$, $\Delta m_{12}^2 = 7.54 \times 10^{-5}$ eV$^2$, $\Delta m_{13}^2 = 2.47\times 10^{-3}$ eV$^2$, and $\delta = 0$, i.e., no ``standard'' $T$-invariance violation. The green curve corresponds to $P_{e\mu}$ while the purple curve corresponds to $P_{\mu e}$. If, instead, all non-zero NSI are real ($\epsilon_{e\mu} = 0.1$, $\epsilon_{e\tau} = 0.1$, $\epsilon_{\mu\tau} = 0.1$), $P_{e\mu}=P_{\mu e}$, the grey curve. The dashed line corresponds to the pure three-neutrino oscillation probabilities assuming no $T$-invariance violation (all $\epsilon_{\alpha\beta}=0$, $\delta=0$). }
\label{fig:TViolation}
\end{figure}

An interesting feature of off-diagonal NSI is that they can mediate $CP$-invariance violating effects (\cite{GonzalezGarcia:2001mp,Winter:2008eg}. These are, however, difficult to identify by comparing the two $CP$-conjugated process $P_{\alpha\beta}$ and $P_{\bar{\alpha}\bar{\beta}}$ because of the $CPT$-violating nature of the matter effects. For illustrative purposes, it is more useful to look at the $T$-conjugated channels $P_{\alpha\beta}$ and $P_{\beta\alpha}$. In the absence of fundamental $CP$-violating parameters, $P_{\alpha\beta}= P_{\beta\alpha}$ even if the neutrinos propagate in matter, as long as the matter-profile is constant (or symmetric upon exchange of the source and the detector). For more details on $T$-violation see, for example, Refs.~\cite{Arafune:1996bt,Koike:1999tb,deGouvea:2000un,Akhmedov:2001kd,Akhmedov:2004ve,Kisslinger:2012qe,Xing:2013uxa}.\footnote{Interesting aside: if there are only two neutrino flavors, $P_{\alpha\beta}=P_{\beta\alpha}$, even if the NSI parameter is complex. This is a direct consequence of the unitarity of the time-evolution of the neutrino state. For more details see, for example, Ref.~\cite{DeGouvea:2001ag}.} Fig.~\ref{fig:TViolation} depicts the $T$-invariance violating effect of the NSI. The oscillation probabilities $P_{\mu e}$ and $P_{e\mu}$ were computed for complex and real NSI parameters, assuming no ``standard'' sources of $T$-invariance violation, i.e.,  $\delta = 0$. We discuss in Sec.~\ref{sec:Sensitivity} how sensitive is the DUNE experiment to these new sources of $T$-invariance (and hence $CP$-invariance) violation. 

\section{Exclusion Capability of DUNE}
\label{sec:Exclusion}
We investigate the sensitivity of the proposed Deep Underground Neutrino Experiment (DUNE)~\cite{Adams:2013qkq} to NSI. We consider that DUNE consists of a $34$ kiloton liquid argon detector and utilizes a $1.2$ MW proton beam to produce neutrino and antineutrino beams from pion decay in flight originating $1300$ km upstream at Fermilab, consistent with the proposal in Ref.~\cite{Adams:2013qkq}. The neutrino energy ranges between $0.5$ and $20$ GeV and the flux is largest around $3.0$ GeV. In the following analyses, we simulate six years of data collection: 3 years each with the neutrino and antineutrino beams. Unless otherwise mentioned, for concreteness, we consider that the true value of $\delta$, the $CP$-violating phase, is $\pi/3$ and restrict our analyses to the normal neutrino mass hierarchy, i.e.\ $\Delta m_{13}^2 > 0$.

\begin{figure}[!htbp]
	\begin{center}
	\includegraphics[width=0.9\linewidth]{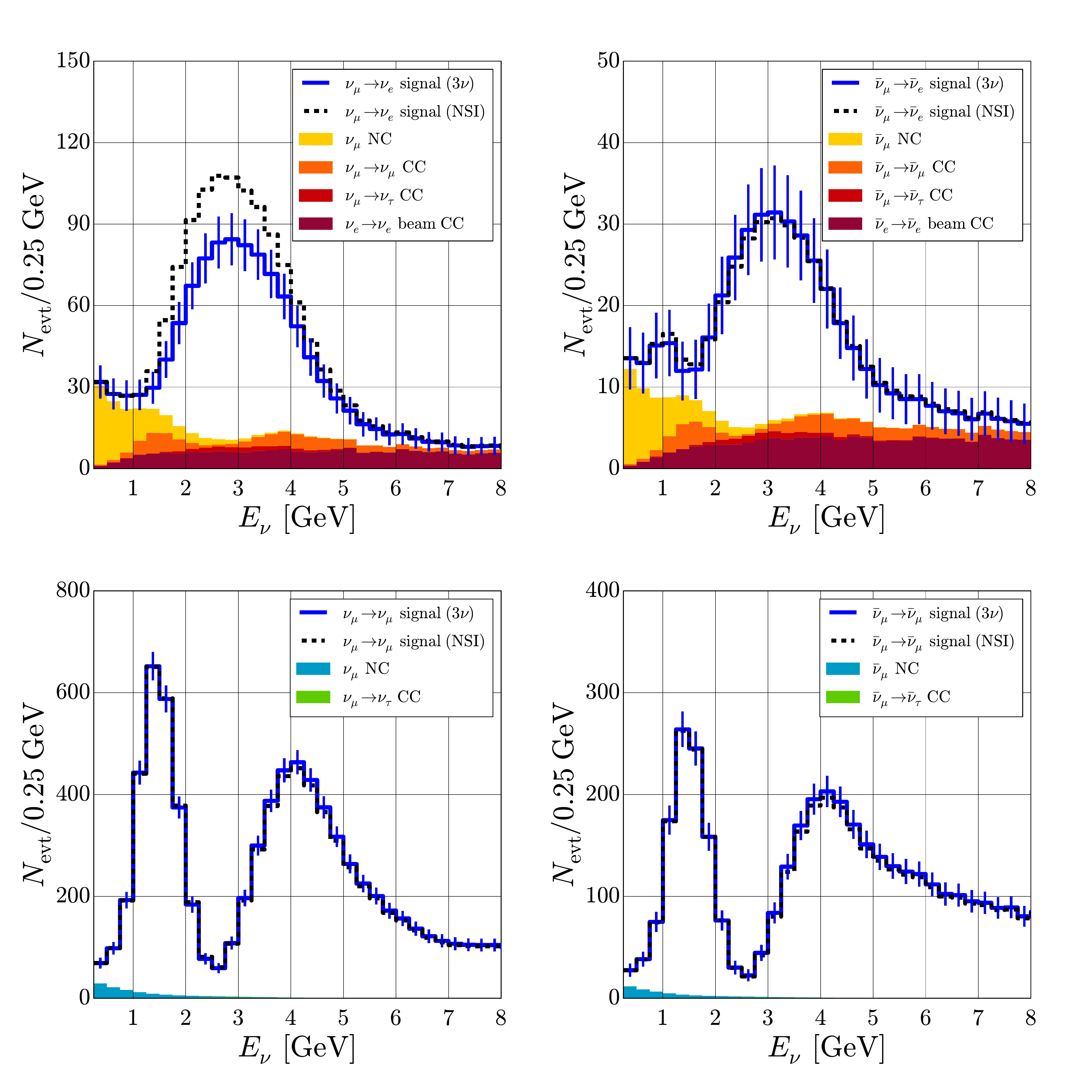}	
	\caption{Expected signal and background	yields for six years (3y $\nu\ +$ 3y $\bar{\nu}$) of data collection at DUNE, using fluxes projected by Ref.~\cite{Adams:2013qkq}, for a 34 kiloton detector, and a 1.2 MW proton beam on target. The top row shows appearance channel yields for neutrino (left) and antineutrino (right) beams, while the bottom row shows disappearance channel yields. `$3\nu$' corresponds to the standard paradigm, with $\sin^2\theta_{12} = 0.308,\ \sin^2\theta_{13} = 0.0234,\ \sin^2\theta_{23} = 0.437,\ \Delta m_{12}^2 = 7.54 \times 10^{-5}$ eV$^2$, $\Delta m_{13}^2 = +2.47\times 10^{-3}$ eV$^2$, $\delta = \pi/3$. `NSI' further assumes $\epsilon_{ee} = 0.5$ and $\epsilon_{\mu\tau} = 0.2e^{-i\pi/2}$. Statistical uncertainties are shown as vertical bars. Backgrounds are defined in the text and are assumed to be identical for the standard paradigm and the NSI scenario.}
	\label{fig:4Pane}
	\end{center}
\end{figure}
We calculate expected event yields for the appearance ($P_{\mu e},P_{\bar{\mu}\bar{e}}$) and disappearance ($P_{\mu \mu},P_{\bar{\mu}\bar{\mu}}$) channels for neutrino and antineutrino beam modes using the oscillation probabilities discussed above, the projected DUNE neutrino fluxes from Ref.~\cite{Adams:2013qkq}, and the neutrino-nucleon cross-sections tabulated in Ref.~\cite{Formaggio:2013kya}. Fig.~\ref{fig:4Pane} depicts these yields for two scenarios. In all four panels the blue line corresponds to a three-neutrino scheme with three-neutrino oscillation parameters as listed in Sec.~\ref{sec:Formalism} (in agreement with Ref.~\cite{Agashe:2014kda}), and  $\delta = \pi/3$. The dashed line corresponds to a nonzero NSI scenario, $\epsilon_{ee} = 0.5$ and $\epsilon_{\mu\tau} = 0.2e^{-i\pi/2}$ (all others are set to zero). The four dominant backgrounds are consequences of muon-type neutrino neutral-current scattering (``$\nu_\mu$ NC''), tau-type neutrino charged-current scattering (``$\nu_\mu \rightarrow \nu_\tau$ CC''), muon-type neutrino charged-current scattering (``$\nu_\mu \rightarrow \nu_\mu$ CC''), and beam electron-type neutrino charged-current scattering (``$\nu_e\rightarrow \nu_e$ beam CC''), as depicted in Fig.~\ref{fig:4Pane}. The rates associated with these backgrounds are taken from Ref.~\cite{Adams:2013qkq}, and event yields have been shown to be consistent with projections by DUNE in Ref.~\cite{Berryman:2015nua}. As in Refs.~\cite{Adams:2013qkq,Berryman:2015nua}, we use 1\% signal and 5\% background normalization uncertainties.

\begin{figure}[!htbp]
\centering
\includegraphics[width=0.8\linewidth]{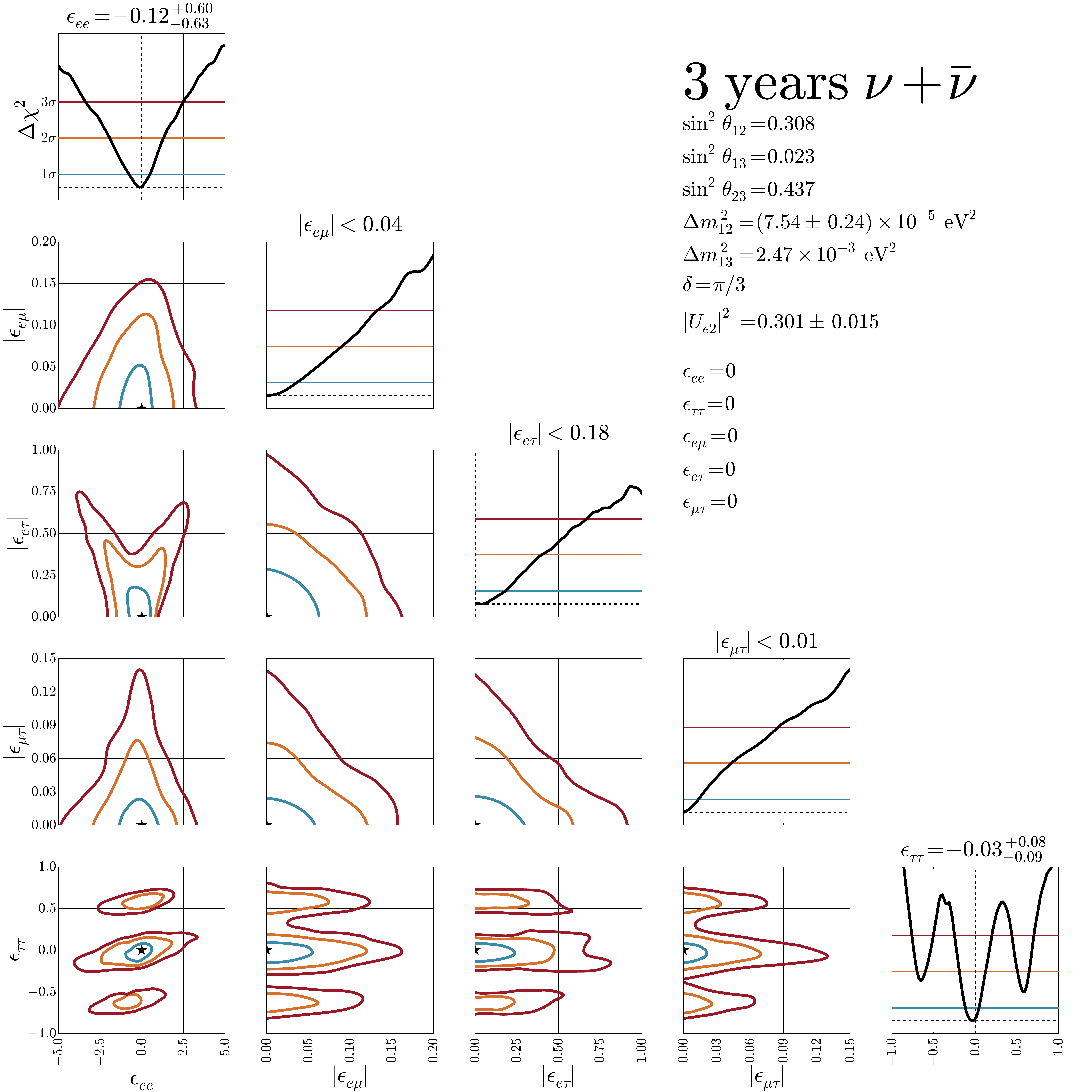}
\caption{Expected exclusion limits at 68.3\% (red), 95\% (orange), and 99\% (blue) CL at DUNE assuming data consistent with the standard paradigm. The $CP$-violating phase $\delta$ is assumed to be $\pi/3$ and the mass hierarchy is normal. Gaussian priors are included on $\Delta m_{12}^2 = (7.54\pm 0.24)\times 10^{-5}$ eV$^2$ and $|U_{e2}|^2 = 0.301\pm 0.015$. See text for details.}
\label{fig:DUNEExclusion}
\end{figure}

If the data collected at DUNE are consistent with the standard paradigm, we can calculate exclusion limits on the parameters $\epsilon_{\alpha\beta}$. We make use of the Markov Chain Monte Carlo program {\sc emcee} to generate probability distributions as a function of several parameters and construct confidence intervals~\cite{ForemanMackey:2012ig}. Fig.~\ref{fig:DUNEExclusion} depicts two-dimensional exclusion limits and one-dimensional reduced $\chi^2$ distributions for the NSI parameters, assuming the generated data to be consistent with the standard paradigm. During the fit, we marginalize over all unseen parameters, including the phases of the off-diagonal NSI parameters and the standard three-neutrino parameters. Additionally, we include Gaussian priors on the solar neutrino parameters: $\Delta m_{12}^2 = (7.54\pm 0.24)\times 10^{-5}$ eV$^2$, $|U_{e2}|^2 = 0.301\pm 0.015$, as discussed in detail in \cite{Berryman:2015nua}. Finally, we choose $\epsilon_{\mu\mu}\equiv 0$. As discussed in the previous section, this is equivalent to reinterpreting $\epsilon_{ee}\to\epsilon_{ee}-\epsilon_{\mu\mu}$ and  $\epsilon_{\tau\tau}\to\epsilon_{\tau\tau}-\epsilon_{\mu\mu}$. Henceforth, we will refer to $\epsilon_{\alpha\alpha}-\epsilon_{\mu\mu}$ as $\epsilon_{\alpha\alpha}$ ($\alpha=e,\tau$) and will no longer discuss $\epsilon_{\mu\mu}$ as an independent parameter. The reason for singling out the $\mu\mu$-component of the NSI is simple: it is the best independently constrained parameter. Depicted in Fig.~\ref{fig:DUNEExclusion}, the DUNE-reach to the diagonal NSI parameters is inferior to the conservative bound on $\epsilon_{\mu\mu}$ in Eq.~(\ref{Bounds}) so the distinction between $\epsilon_{\alpha\alpha}$ and $\epsilon_{\alpha\alpha}-\epsilon_{\mu\mu}$ is not, in practice, significant.  

The expected bounds from DUNE are significantly stronger than those displayed in Eq.~(\ref{Bounds}) for all relevant NSI parameters. The $\sin^2\theta_{23}$ - $\epsilon_{\tau\tau}$ degeneracy discussed in connection with Eq.~(\ref{Diag}) in the previous section manifests itself in the form of the local minima for the reduced $\chi^2(\epsilon_{\tau\tau})$ in Fig.~\ref{fig:DUNEExclusion} (fifth row, far right). In agreement with the discussion surrounding Eq.~(\ref{OffDiag}), we find that the larger allowed values of $|\epsilon_{\mu\tau}|$ are associated with  $\mathrm{arg}(\epsilon_{\mu\tau}) = \pi/2$ or $3\pi/2$, which is responsible for the ``flat shape'' for the for the reduced $\chi^2(|\epsilon_{\mu\tau}|)$ in Fig.~\ref{fig:DUNEExclusion} (fourth row, far right). The $|\epsilon_{\mu\tau}|$ - $\mathrm{arg}(\epsilon_{\mu\tau})$ plane is not depicted in Fig.~\ref{fig:DUNEExclusion}, but we have verified that $\mathrm{arg}(\epsilon_{\mu\tau}) = \pi/2$ or $3\pi/2$ correlates with large allowed values of $|\epsilon_{\mu\tau}|$. This type of behavior will be discussed in more detail in Sec.~\ref{sec:Sensitivity}.

Previous studies, particularly those of atmospheric oscillations, observed degeneracies in the $\epsilon_{e\tau}$ - $\epsilon_{\tau\tau}$ plane for fixed values of $\epsilon_{ee}$~\cite{Friedland:2004ah,Yasuda:2010hw}. We find that DUNE's increased sensitivity eliminates this degeneracy for the most part, and only modest hints of correlations can be seen in the $\epsilon_{ee}$ - $|\epsilon_{e\tau}|$ plane in Fig.~\ref{fig:DUNEExclusion} (third row, far left).


\section{NSI Discovery Potential of DUNE}
\label{sec:Sensitivity}

Here, we assume that the data collected at DUNE is consistent with nonzero NSI. In order to explore different scenarios and discuss the discovery potential of DUNE more quantitatively, we consider three concrete, qualitatively different cases, tabulated in Table~\ref{CaseTable}. In all cases the values of the selected  $\epsilon_{\alpha\beta}$ are within the bounds in Eq.~(\ref{Bounds}) but outside the expected allowed regions in Fig.~\ref{fig:DUNEExclusion}.

\begin{table}[htbp]
\centering
\begin{tabular}{| c | c | c | c | c | c | c | }
\hline
 & $\epsilon_{ee}$ & $\epsilon_{e\mu}$ & $\epsilon_{e\tau}$ & $\epsilon_{\mu\mu}^{\star}$ & $\epsilon_{\mu\tau}$ & $\epsilon_{\tau\tau}$ \\
 \hline
 \textbf{Case 1} & $0$ & $0.15e^{i\pi/3}$ & $0.3e^{-i\pi/4}$ & 0 & $0.05$ & 0\\
 \hline
 \textbf{Case 2} & $-1.0$ & $0$ & $0$ & $0$ & 0 & $0.3$\\
 \hline
 \textbf{Case 3} & $0.5$ & $0$ & $0.5e^{i\pi/3}$ & $0$ & $0$ & $-0.3$\\
 \hline
\end{tabular}
\caption{Input values of the new physics parameters for the three NSI scenarios under consideration. The star symbol is a reminder that,  as discussed in the text, we can choose $\epsilon_{\mu\mu}\equiv 0$ and reinterpret the other diagonal NSI parameters.}
\label{CaseTable}
\end{table}

In Case 1, we assume that all NSI are strictly off-diagonal (in the neutrino flavor space). We also allow large relative phases between the different $\epsilon_{\alpha\beta}$. In Case 2, we assume all NSI are diagonal in flavor space, and, as discussed earlier, fix $\epsilon_{\mu\mu}\equiv0$. Finally, in Case 3, we assume that the new physics lies within the $e-\tau$ sector in such a way that $\epsilon_{\alpha\mu} = 0$ for all $\alpha=e,\mu,\tau$. This case is partially motivated by the fact that $\epsilon_{\alpha\mu}$ are, currently, the best constrained NSI parameters. 

As in the previous section, for all simulations, we consider three years each of neutrino and antineutrino mode data collection, with a $34$ kt detector and $1.2$ MW beam. Three-neutrino parameters are consistent with Ref.~\cite{Agashe:2014kda}. The data are analyzed using a $\chi^2$ function, and we utilize {\sc emcee} to generate parameter likelihood distributions in one and two dimensions. When defining a best-fit point to simulated data, we consider the overall minimum of the $\chi^2$ distribution, as opposed to the set of minima of the marginalized one-dimensional parameter likelihoods. We do not recalculate background yields for the different hypotheses as the 5\% background normalization uncertainty renders any discrepancies negligible.

\subsection{Compatibility with the Standard Paradigm}
\label{subsec:3NuFit}

The first question one needs to address is whether DUNE can successfully diagnose, if NSI are real, that the standard paradigm is not acceptable. To pursue this question, we simulate data assuming the three cases discussed above (Table~\ref{CaseTable}), but attempt to fit to the data assuming the standard paradigm. Following the philosophy of Ref.~\cite{Berryman:2015nua}, we do not attempt to combine DUNE data with existing data or data from any other concurrent future experiments, except when it comes to ``solar'' data. As in the previous section, we include our current knowledge of the solar parameters $\Delta m^2_{12}$ and $\sin^2\theta_{12}$ by imposing Gaussian priors on $\Delta m_{12}^2 = (7.54\pm 0.24)\times 10^{-5}$ eV$^2$, $|U_{e2}|^2 = 0.301\pm 0.015$ in our analysis, unless otherwise noted. We point readers to Ref.~\cite{Berryman:2015nua} for more details. We comment on combining DUNE data with those of other future experiments in the Conclusions. 

For the three cases we explore, the standard paradigm provides a poor fit to the simulated data. For Cases 1, we find, for the best-fit point, $\chi^2_\text{min} / $ dof = $261/114$, or a $7.4\sigma$ discrepancy. Case 2 yields a similarly poor fit, $\chi^2_\text{min} / $ dof = $259/114$, or roughly a $7.3\sigma$ discrepancy. For Case 3 we find the fit is mediocre, and not as extreme as the previous cases: $\chi^2_\text{min} /$ dof $= 142/114$, a $2.1\sigma$ discrepancy. We also explored the consequences of removing the Gaussian priors on the solar parameters, and the DUNE-only fits improve significantly: Case 1, $\chi^2_\text{min} / $ dof $= 204/114$, a $5.1\sigma$ discrepancy, Case 2, $\chi^2_\text{min} /$ dof $= 166/114$, a $3.3\sigma$ discrepancy, and Case 3, $\chi^2_\text{min} /$ dof $=139/114$, a $1.9\sigma$ discrepancy. For all the three cases, the preferred value of $\Delta m^2_{12}$ is too large ($\Delta m^2_{12}>10^{-4}$~eV$^2$), while the preferred values for the solar angle are also significantly outside the currently allowed range. The quality of the different fits to the different cases are tabulated in Table~\ref{FitSummaryTable}.

\subsection{Measuring the NSI Parameters}
\label{subsec:NSIFit}

Once it is established that the standard paradigm is ruled out, one can proceed to try to explain the data using a new physics model. Here we discuss the results of fitting the NSI-consistent DUNE data with the NSI hypothesis, and show how well one can measure the new physics parameters $\epsilon_{\alpha\beta}$. As discussed in Sec.~\ref{sec:Exclusion}, we choose $\epsilon_{\mu\mu}\equiv 0$.

\begin{figure}[ht]
\centering
\includegraphics[width=0.9\linewidth]{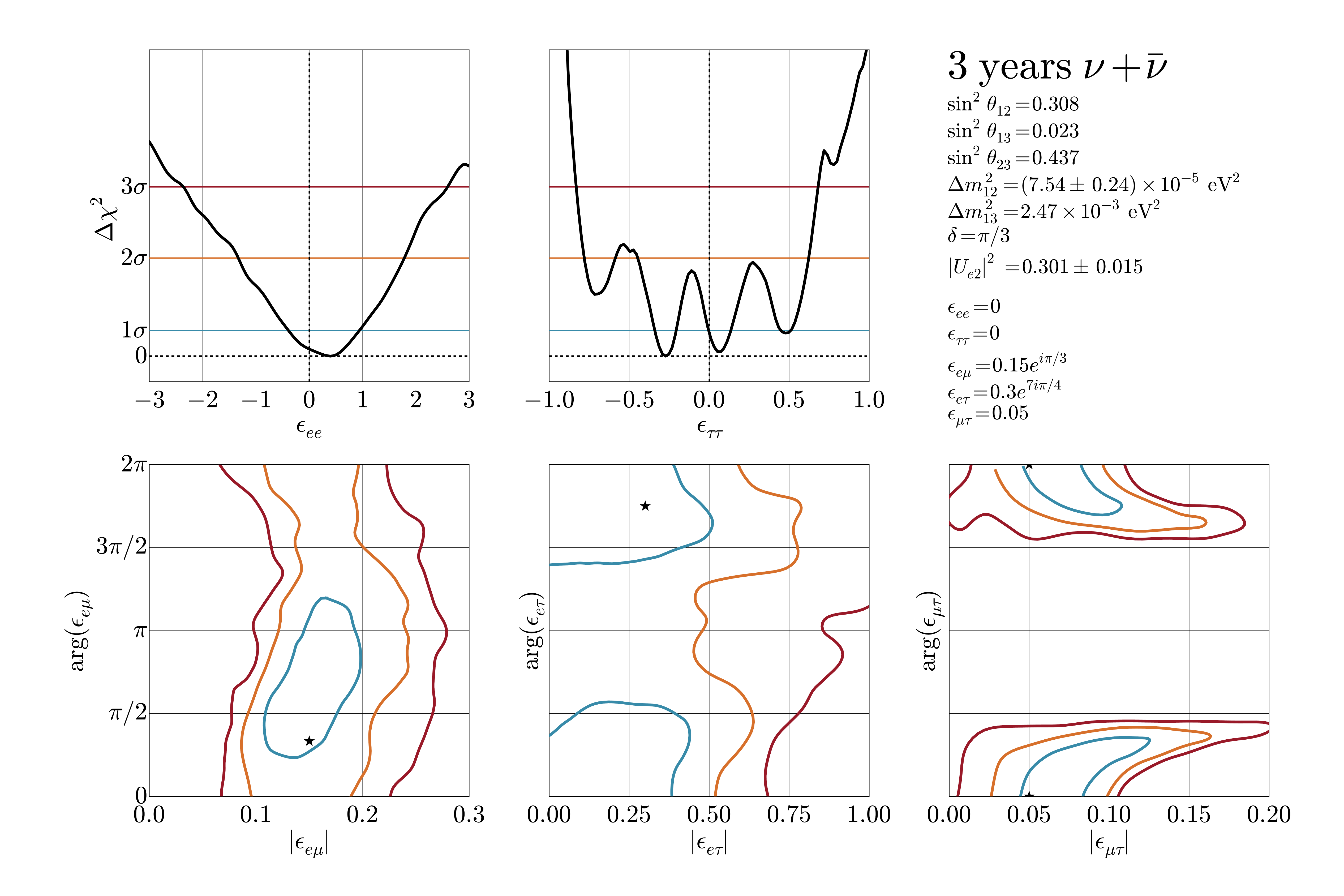}
\caption{Expected measurement sensitivity for real (top) and complex (bottom) parameters at DUNE, assuming Case 1 (see Table~\ref{CaseTable}). Top row: reduced $\chi^2$, with horizontal lines indicating the $1$, $2$, and $3\sigma$ allowed ranges. Bottom row: sensitivity contours at 68.3\% (blue), 95\% (orange), and 99\% (red) CL. The input three-neutrino oscillation parameters are consistent with Ref.~\cite{Agashe:2014kda} and are marginalized over. Gaussian priors are included on the values of $\Delta m_{12}^2$ and $|U_{e2}|^2$. See text for details.}
\label{fig:C1Sensitivity}
\end{figure}

Fig.~\ref{fig:C1Sensitivity} depicts calculated sensitivity contours to the NSI parameters when data generated are consistent with Case 1 in Table~\ref{CaseTable}. During the data-analysis, while quoting the allowed values of the different $\epsilon_{\alpha\beta}$, we marginalize over all other oscillation parameters, including the standard three-neutrino oscillation parameters. Here, DUNE can exclude $|\epsilon_{e\mu}| =0$ and $|\epsilon_{\mu\tau}| =0$ at almost the 99\% CL, and can constrain the diagonal $|\epsilon_{\alpha\alpha}|\lesssim 1$ at the one-sigma ($\alpha = e$) or three-sigma ($\alpha=\tau$) level. If Case 1 happens to be consistent with DUNE data, the experiment will also be able to provide hints, at the one sigma level, that $\epsilon_{e\mu}$ is complex and that, hence, there is a source of $CP$-invariance violation in the lepton sector other than $\delta$. 

\begin{figure}[ht]
\centering
\includegraphics[width=0.9\linewidth]{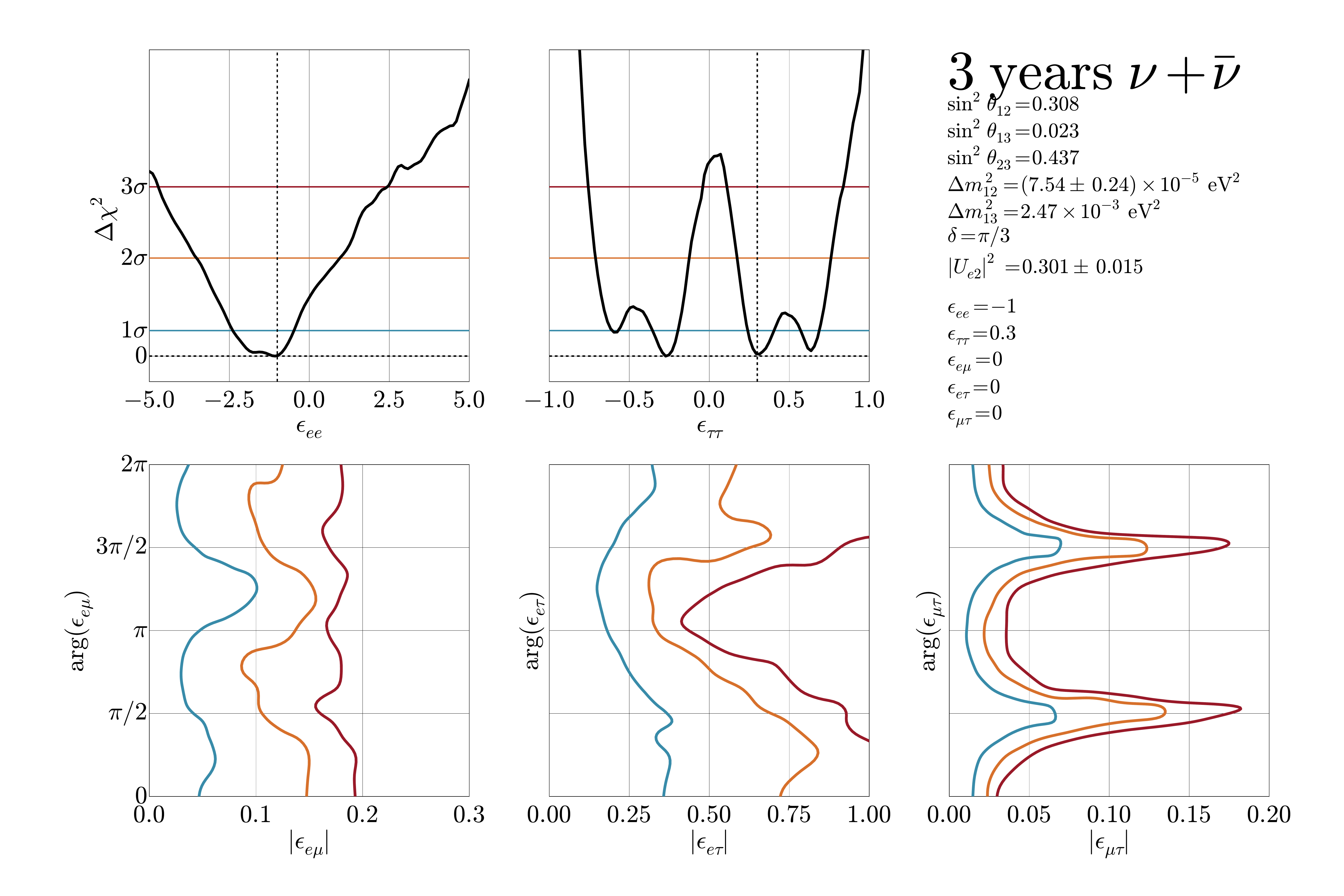}
\caption{Expected measurement sensitivity for real (top) and complex (bottom) parameters at DUNE, assuming Case 2 (see Table~\ref{CaseTable}). Top row: reduced $\chi^2$, with horizontal lines indicating the $1$, $2$, and $3\sigma$ allowed ranges. Bottom row: sensitivity contours at 68.3\% (blue), 95\% (orange), and 99\% (red) CL. The input three-neutrino oscillation parameters are consistent with Ref.~\cite{Agashe:2014kda} and are marginalized over. Gaussian priors are included on the values of $\Delta m_{12}^2$ and $|U_{e2}|^2$. See text for details.}
\label{fig:C2Sensitivity}
\end{figure}

In Case 2, only the diagonal NSI parameters are non-zero: $\epsilon_{ee} = -1.0$, $\epsilon_{\tau\tau} = 0.3$. Their measurements are depicted in Fig. \ref{fig:C2Sensitivity}. Here, $\epsilon_{ee} = 0$ is excluded at between 68.3\% and 95\% CL, and $\epsilon_{\tau\tau} = 0$ is excluded at 99\% CL. The local minima of $\chi^2(\epsilon_{\tau\tau})$ in Fig. \ref{fig:C2Sensitivity} are mostly a manifestation of the degeneracy between $\epsilon_{\tau\tau}$ and $\sin^2\theta_{23}$, discussed earlier in the text surrounding Eq.~(\ref{Diag}). The $|\epsilon_{\mu\tau}|$ - $\mathrm{arg}(\epsilon_{\mu\tau})$ plane (Fig. \ref{fig:C2Sensitivity} bottom-right) also illustrates the effect discussed around Eq.~(\ref{OffDiag}) -- the sensitivity to nonzero $|\epsilon_{\mu\tau}|$ is least when $\mathrm{arg}(\epsilon_{\mu\tau}) = \pi/2$ or $3\pi/2$. 

\begin{figure}[ht]
\centering
\includegraphics[width=0.9\linewidth]{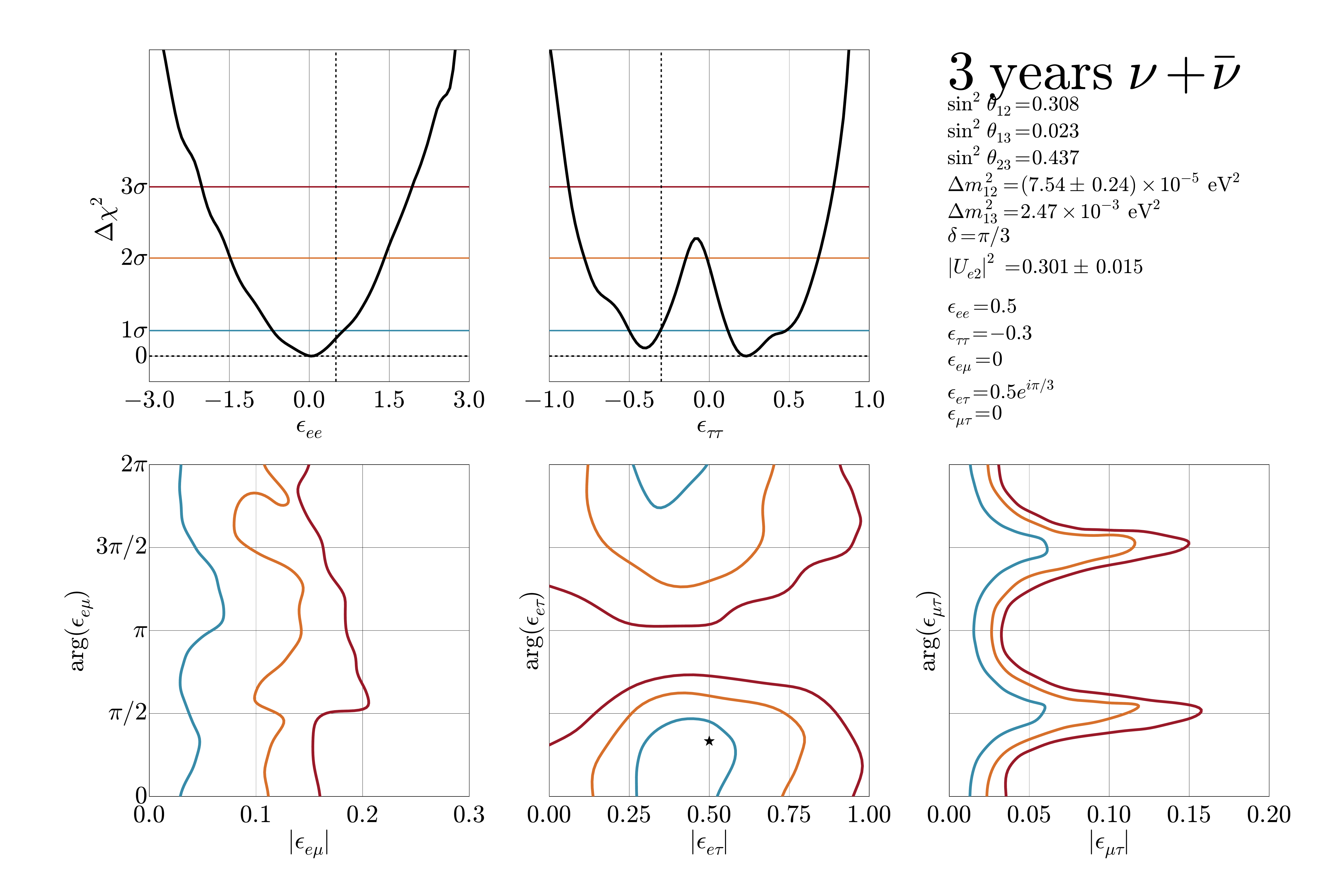}
\caption{Expected measurement sensitivity for real (top) and complex (bottom) parameters at DUNE, assuming Case 3 (see Table~\ref{CaseTable}). Top row: reduced $\chi^2$, with horizontal lines indicating the $1$, $2$, and $3\sigma$ allowed ranges. Bottom row: sensitivity contours at 68.3\% (blue), 95\% (orange), and 99\% (red) CL. The input three-neutrino oscillation parameters are consistent with Ref.~\cite{Agashe:2014kda} and are marginalized over. Gaussian priors are included on the values of $\Delta m_{12}^2$ and $|U_{e2}|^2$.  See text for details.}
\label{fig:C3Sensitivity}
\end{figure}

In Case 3, we allow both diagonal and off-diagonal NSI to be non-zero, but imposed that all new physics parameters involving the muon-flavor to be zero: $\epsilon_{ee} = 0.5$, $\epsilon_{\tau\tau} = -0.3$, $\epsilon_{e\tau} = 0.5e^{i\pi/3}$. Their measurements are depicted in Fig.~\ref{fig:C3Sensitivity}. Here, we see good sensitivity to the non-zero parameters, except $\epsilon_{ee}$, where the input value is too close to zero to distinguish. $\epsilon_{\tau\tau} = 0$ is excluded at roughly 95\% CL, and we see the previously discussed degeneracy between $\epsilon_{\tau\tau}$ and $\sin^2\theta_{23}$. Additionally, $|\epsilon_{e\tau}|=0$ is excluded at over 95\% CL, and we see that part of the phase space for arg$(\epsilon_{e\tau})$ is ruled out at 99\% CL. Lastly, we see again the reduced sensitivity to $\epsilon_{\mu\tau}$ when it is purely imaginary.

\subsection{Compatibility with a Sterile Neutrino?}

While we have established that DUNE data are inconsistent with the standard paradigm if they happen to be consistent with one of the three cases tabulated in Table~\ref{CaseTable}, it remains to address whether the data are consistent with other new-physics scenarios. In other words, we would like to diagnose whether DUNE can not only establish that the standard paradigm is wrong but also want to understand whether DUNE can distinguish different potential new physics scenarios. Here we compare NSI hypothesis with the hypothesis that there are new neutrino states.\footnote{This is, of course, one of many possibilities. The authors of Ref.~\cite{Meloni:2009cg}, for example, compared NSI effects with non-unitarity effects at a Neutrino Factory.}

We fit the different simulated data sets introduced earlier in this section (Table~\ref{CaseTable}) to a four-neutrino hypothesis. As in Ref.~\cite{Berryman:2015nua}, we assume $m_4 > m_1$ (but not necessarily $m_4 > m_3$) and include the two new relevant mixing angles, $\phi_{14}$ and $\phi_{24}$, and one new $CP$-violating phase. We refer readers to Ref.~\cite{Berryman:2015nua} for more details, including the definition of the new physics parameters. 

The NSI and sterile-neutrino hypotheses are qualitatively different. They mediate different energy dependent effects, for example. NSI don't lead to new oscillatory behavior but, instead, (roughly speaking) grow in importance as the energy grows. On the other hand, the sterile neutrino hypothesis is quite versatile. As one varies the magnitude of the new mass-squared difference, inside a fixed $L/E_{\nu}$-range, the new oscillation effects vary between fast, averaged-out effects (for a large new mass-squared difference $\Delta m^2_{\rm new}$), new oscillatory features (for $\Delta m^2_{\rm new}L/E_{\nu}\sim 1$), to effects that grow slowly with growing $L/E_{\nu}$ and ultimately mimic the violation of the unitarity of the PMNS matrix (for $\Delta m^2_{\rm new}L/E_{\nu}\ll  1$). Overall, we expect that a four-neutrino interpretation of NSI-consistent data will fare better than a three-neutrino interpretation. If nothing else, there are more free parameters in the four-neutrino hypothesis, and the four neutrino hypothesis includes the three neutrino one.

For the fit to Case 1 data, the four-neutrino fit (including priors from Solar neutrino data) is significantly improved when compared to the three-flavor fit discussed earlier (Sec.~\ref{subsec:3NuFit}): $\chi^2_{\text{min}}/$ dof $= 181/110$, or a $4.2\sigma$ discrepancy, compared to the $7.4\sigma$ discrepancy for the standard-paradigm fit. The fit to Case 2 is significantly improved, from the $7.3\sigma$ discrepancy mentioned in Sec.~\ref{subsec:3NuFit} to $\chi^2_{\text{min}} /$ dof $= 149/110$, or a $2.7\sigma$ discrepancy. The quality of the four-neutrino fit to Case 3 data is $\chi^2_{\text{min}} /$ dof $= 120/110$, or a $1.2\sigma$ discrepancy. This is to be compared to a $2.1\sigma$ discrepancy for the three-neutrino fit. Results of all of the fits are summarized in Table~\ref{FitSummaryTable}.

\begin{table}[htbp]
\centering
\begin{tabular}{| c || c | c | c |}
\hline
Fit & Case 1 & Case 2 & Case 3 \\
\hline
3$\nu$ with Solar Priors & $261/114 \simeq 7.4\sigma$ & $259/114 \simeq 7.3\sigma$ & $142/114 \simeq 2.1\sigma$ \\
\hline
3$\nu$ without Priors & $204/114 \simeq 5.1\sigma$ & $166/114 \simeq 3.3\sigma$ & $139/114 \simeq 1.9\sigma$ \\
\hline
4$\nu$ with Solar Priors & $181/110 \simeq 4.2\sigma$ & $149/110 \simeq 2.7\sigma$ & $120/110 \simeq 1.2\sigma$ \\
\hline
\end{tabular}
\caption{Results of various three- or four-neutrino fits to data generated to be consistent with the cases listed in Table~\ref{CaseTable}. Numbers quoted are for $\chi^2_{\text{min}} /$ dof and the equivalent discrepancy using a $\chi^2$ distribution.}
\label{FitSummaryTable}
\end{table}

\begin{figure}[ht]
\centering
\includegraphics[width=0.7\linewidth]{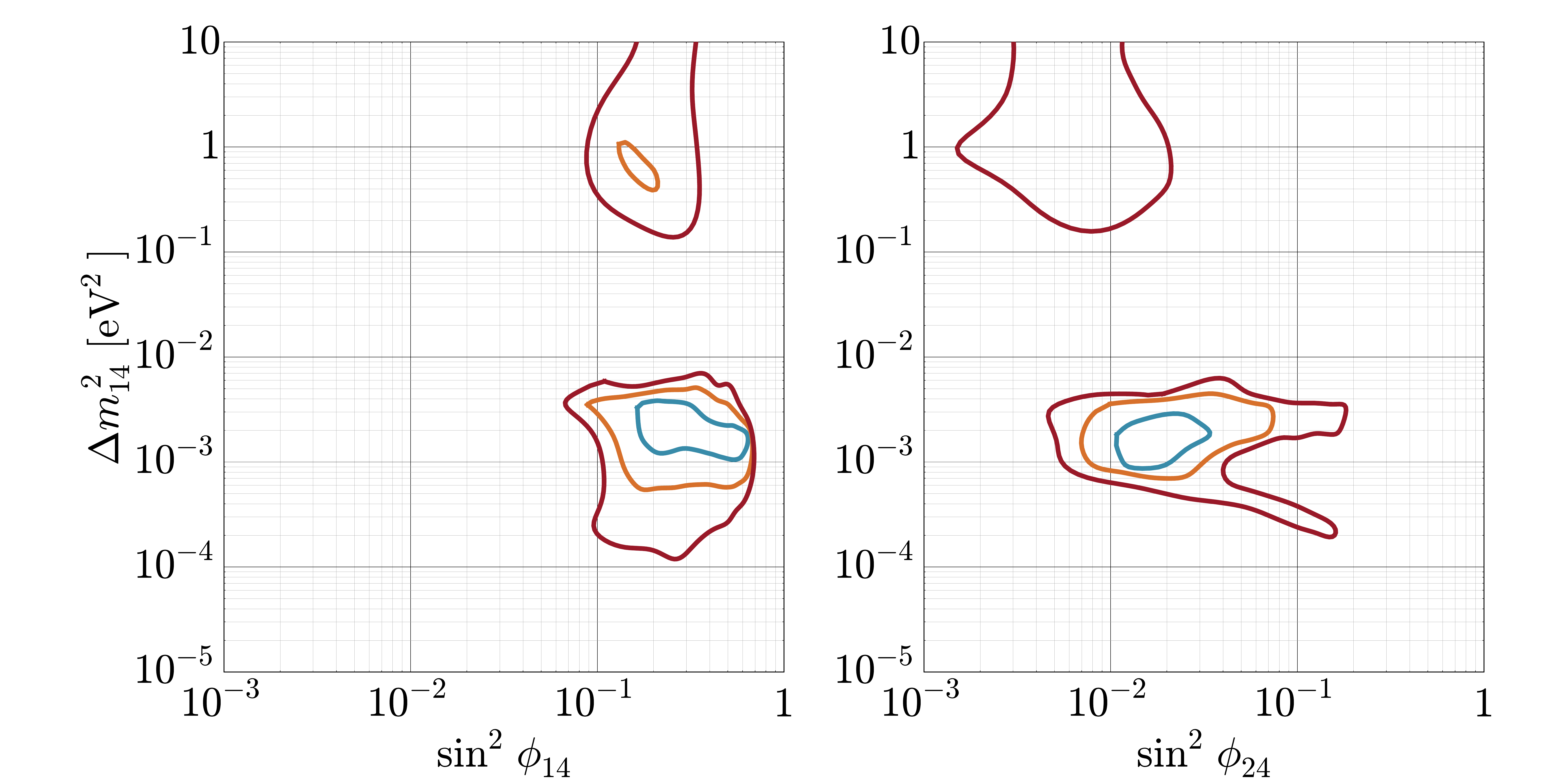}
\caption{Sensitivity contours at 68.3\% (blue), 95\% (orange), and 99\% (red) for a four-neutrino fit to data consistent with Case 2 from Table~\ref{CaseTable}. All unseen parameters are marginalized over, and Gaussian priors are included on the values of $\Delta m_{12}^2$ and $|U_{e2}|^2$. See text for details.}
\label{fig:4NuFit_C2}
\end{figure}

For all three cases, the four-neutrino fit favors large values of $\phi_{14}$ and $\phi_{24}$, and $\Delta m_{14}^2 \simeq \Delta m_{13}^2$\footnote{The four-neutrino fit to case 3 is also compatible with $\Delta m_{14}^2 \simeq 10^{-6} - 10^{-5}$ eV$^2$, a region not yet excluded by existing experiments.}. The reason for this is that NSI effects are energy dependent and, in the case of a four-neutrino scenario, a new observable oscillation frequency requires $\Delta m_{14}^2 \simeq \Delta m_{13}^2$. For illustrative purposes, Fig.~\ref{fig:4NuFit_C2} depicts the allowed region of the four-neutrino parameter space assuming DUNE data is consistent with Case 2 in Table~\ref{CaseTable}. The fit is compatible with $\sin^2\phi_{14}\sim 10^{-1}$ and $\sin^2\phi_{24}\sim 10^{-2}$ and $\Delta m_{14}^2 \sim 10^{-3}$~eV$^2$. Such large values of $\sin^2\phi_{14}$ for this range of $\Delta m^2_{14}$ are, currently, already strongly constrained by the Daya Bay experiment \cite{An:2014bik}.

\section{Conclusions}
\label{sec:Conclusions}

We explored the effects of NSI and how they modify neutrino propagation in the DUNE experimental setup. We find that NSI can significantly modify the data to be collected by the DUNE  experiment as long as the new physics parameters are ${\cal O}(10^{-1})$. This means that if the DUNE data is consistent with the standard paradigm, ${\cal O}(10^{-1})$ NSI effects will be ruled out. Details are depicted in see Fig.~\ref{fig:DUNEExclusion}. On the other hand, if large NSI effects are present, DUNE will be able to not only rule out the standard paradigm\footnote{Strictly speaking, as discussed in Sec.~\ref{subsec:3NuFit}, this conclusion requires that one also include currently available information from experiments sensitive to the solar parameters $\Delta m^2_{12}$ and $\sin^2\theta_{12}$.} but also measure the new physics parameters. Several concrete hypothetical scenarios were considered, and the quantitative results are depicted in Figs.~\ref{fig:C1Sensitivity}, \ref{fig:C2Sensitivity}, \ref{fig:C3Sensitivity}. In particular, the figures reveal that, in some cases, DUNE is sensitive to new sources of $CP$-invariance violation. 

We also explored whether DUNE data can be used to distinguish different types of new physics beyond nonzero neutrino masses. In more detail, we asked, simulating different data sets consistent with the three cases tabulated in Table~\ref{CaseTable}, whether an analysis assuming a four-neutrino hypothesis could yield a good fit. The answer depends on the data set. In some cases a reasonable fit was obtained while in the other cases the four-neutrino fit was poor. For the cases where a reasonable fit is obtained, we are forced to conclude that DUNE data by itself is powerless to significantly distinguish NSI from a sterile neutrino. If this is the case, data from other sources will be required to break the degeneracy among qualitatively different new physics models. 

By the time DUNE starts collecting data, different neutrino oscillation data sets, from a variety of sources, will be available. It is also possible that the HyperK detector will be online and collecting data concurrently with DUNE. Given the fact that both proposed experiments have, for example, similar reach to  $CP$-invariance violating effects that might be present in the standard paradigm, we qualitatively discuss how combined DUNE and HyperK data can be used to disentangle different new physics scenarios, concentrating on NSI versus a sterile neutrino.  

\begin{figure}
\centering
\includegraphics[width=0.6\linewidth]{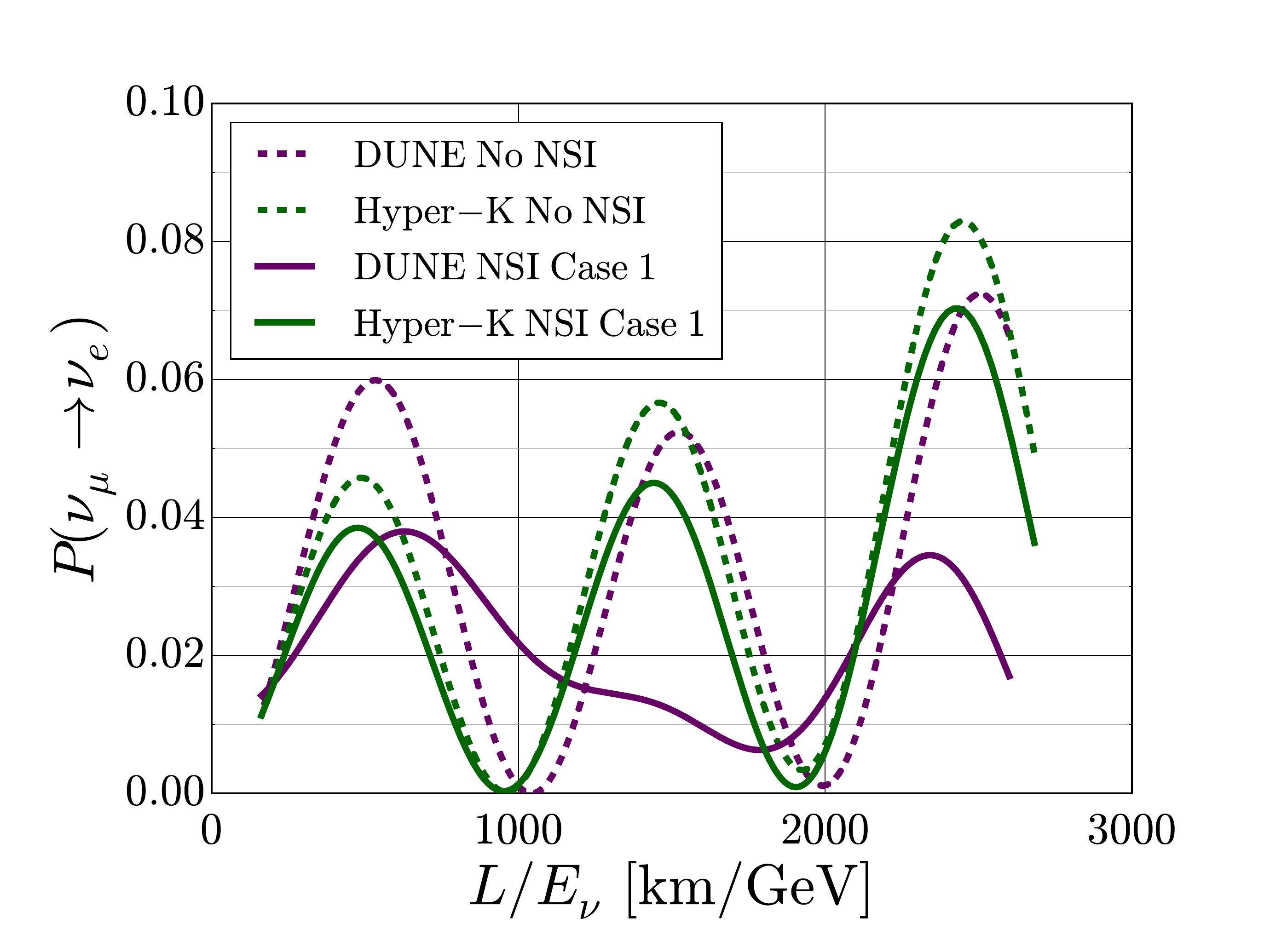}
\caption{Oscillation probabilities for three-neutrino (dashed) and NSI (solid) hypotheses as a function of $L/E_\nu$, the baseline length divided by neutrino energy, for the DUNE (purple) and HyperK (green) experiments. Here, $\delta = 0$ and the three-neutrino parameters used are consistent with Ref.~\cite{Agashe:2014kda}.}
\label{DUNEvsHK}
\end{figure}

The key different between the two experiments, when one compares their capabilities as long-baseline beam experiments, is the fact that they operate at very different baselines ($1300$ km and $295$ km, respectively) but similar values of $L/E_{\nu}$.\footnote{For the sake of discussion we will not take into account that DUNE covers a wider range of $L/E_{\nu}$ values.} Sterile neutrino effects -- discounting ordinary matter effects -- scale like $L/E_{\nu}$ and hence are expected to modify both DUNE and HyperK data in the same way. NSI, on the other hand, are not functions of $L/E_{\nu}$. As is well-known, matter effects are stronger at DUNE so NSI will affect DUNE data more than HyperK data. For illustrative purposes, Fig.~\ref{DUNEvsHK} depicts $P_{\mu e}$ oscillation probabilities in the standard paradigm and in the presence of NSI (assuming Case 1 in Table~\ref{CaseTable}) as a function of $L/E_{\nu}$ for the DUNE and HyperK baselines. It is clear that, assuming NSI exist, the oscillation probabilities at DUNE and HyperK are quite different. 

HyperK, along with several other experiments including the Precision IceCube Next Generation Upgrade (PINGU) \cite{Aartsen:2014oha} and the Indian Neutrino Observatory (INO), will collect large atmospheric neutrino data sets. Atmospheric neutrinos have access to a much broader range of $L/E_{\nu}$ values and matter densities. The comparison of DUNE data with future (and current) atmospheric neutirno data is also expected to help resolve the nature of the new physics that might lie beyond the standard paradigm.

We conclude by commenting on the status of NSI as a realistic (or at least plausible) model for new physics in the neutrino sector or, equivalently, whether future data from the DUNE experiment can significantly add to what is already known about well-defined new physics scenarios that might manifest themselves as NSI. The fact that we consider the effective Lagrangian Eq.~(\ref{eq:L}) at energy scales below that of electroweak symmetry breaking invites one to ask whether there are no ``tree-level'' charged-lepton NSI effects mediated by the same heavy physics. If this is the case, constraints on NSI are expected to be very strong (see, for example, \cite{Bergmann:1999pk,Biggio:2009kv,Ohlsson:2012kf,Miranda:2015dra}). The inclusion of bounds from the charged-lepton sector is, however, model dependent. Nonetheless, it is fair to say that avoiding these bounds all together is very tricky. ``Loop-level'' effects still provide very stringent constraints  \cite{Biggio:2009nt,Bellazzini:2010gn} which can only be avoided, from a model building point of view, in rather convoluted ways \cite{Gavela:2008ra}.\footnote{This is true, at least, if one assumes the new physics is heavy. For other approaches see, for example, Ref.~\cite{Farzan:2015doa}. ``Light mediators'' also allow one to avoid constraints from collider experiments (See, for example, \cite{Friedland:2011za,Franzosi:2015wha}).} Nonetheless, we find that NSI still provide a well-defined scenario that modifies neutrino propagation in a way that is calculable and nontrivial. It also serves -- in the very least -- as an excellent straw man for gauging how sensitive different neutrino experiments are to new phenomena, provides a means of comparing the reach of different proposals, and allows one to discuss the wisdom and power of combining data from different types of neutrino experiments.  

{\bf Note Added:} A couple of days after our manuscript first appeared on the preprint arXiv, Ref.~\cite{Coloma:2015kiu}, which also investigates NSI effects at DUNE, also became publicly available. The results discussed there -- as far as overlapping questions are concerned -- are consistent with the ones discussed here.

\begin{acknowledgments}
We are grateful for useful discussions with Jeff Berryman, Pilar Coloma, and Andrew Kobach. This work is supported in part by the DOE grant \#DE-FG02-91ER40684.  
\end{acknowledgments}

\bibliography{NSIBib}{}

\end{document}